\documentclass[conference]{IEEEtran}
\IEEEoverridecommandlockouts

\usepackage{cite}
\usepackage{amsmath,amssymb,amsfonts}
\usepackage{algorithmic}
\usepackage{multirow}
\usepackage{booktabs}
\usepackage{graphicx}
\usepackage{threeparttable}
\usepackage[table,xcdraw]{xcolor}
\usepackage{textcomp}
\usepackage{xcolor}
\usepackage{xspace}
\usepackage{enumitem}
\usepackage{acronym}
\usepackage[binary-units=true, per-mode=symbol]{siunitx}
\usepackage[a4paper, total={184mm,239mm}]{geometry}
\def\BibTeX{{\rm B\kern-.05em{\sc i\kern-.025em b}\kern-.08em
    T\kern-.1667em\lower.7ex\hbox{E}\kern-.125emX}}
\usepackage[shortcuts]{extdash}
\usepackage{fixfoot}
\usepackage[hyphens]{url}
\usepackage[hidelinks]{hyperref}
\usepackage{balance}
\usepackage{textcomp}
\usepackage{multicol}
\hypersetup{breaklinks=true}
\usepackage{soul}

\makeatletter
\let\MYcaption\@makecaption
\makeatother

\usepackage[font=footnotesize,position=b]{subcaption}

\makeatletter
\let\@makecaption\MYcaption
\makeatother

\acrodef{SoA}{state-of-the-art}
\acrodef{ISA}{instruction set architecture}
\acrodef{TCDM}{tightly-coupled data memory}
\acrodef{FP}{floating-point}
\acrodef{FPU}{floating-point unit}
\acrodef{GPU}{graphics processing unit}
\acrodef{FMA}{fused multiply-add}
\acrodef{FP}{floating-point}
\acrodef{ExFMA}{expanding \ac{FMA}}
\acrodef{PE}{processing element}
\acrodef{NN}{neural network}
\acrodef{HW}{hardware}
\acrodef{ExSdotp}{expanding sum-of-dot-product}
\acrodef{Vsum}{vector inner sum}
\acrodef{MAC}{Multiply Accumulate}
\acrodef{ExVsum}{expanding vector inner sum}
\acrodef{FREP}{floating-point repetition instruction}
\acrodef{SSR}{stream semantic register}
\acrodef{TF32}{TensorFloat-32}
\acrodef{GEMM}{general-matrix-multiplication}
\acrodef{CSR}{control and status register}
\acrodef{FCSR}{floating-point control and status register}

\newcommand{\riscv}{RISC\=/V\xspace}
\newcommand{\etal}{\textit{et al.}\xspace}
\newcommand{\figref}[1]{Fig.~\ref{#1}}
\newcommand{\tabref}[1]{Table~\ref{#1}}
\newcommand{\secref}[1]{Section~\ref{#1}}

\DeclareSIUnit\flops{FLOPS}
\DeclareSIUnit\flop{FLOP}
\DeclareSIUnit\cycle{cycle}
\DeclareSIUnit\cycles{cycles}
\DeclareSIUnit\GE{GE}

\renewcommand{\baselinestretch}{0.96}

\usepackage{eso-pic}

\begin{document}
\AddToShipoutPictureBG*{%
  \AtPageUpperLeft{%
    \hspace{\paperwidth}%
    \raisebox{-\baselineskip}{%
      \makebox[-20pt][r]{\footnotesize{This work has been submitted to the IEEE for possible publication. Copyright may be transferred without notice, after which this version may no longer be accessible.}}
}}}%

\title{MiniFloat-NN and ExSdotp: An ISA Extension and a Modular Open Hardware Unit for Low-Precision Training on  RISC-V cores
\thanks{Supported in part by the European Union's H2020 Fractal \#877056 and European PILOT \#101034126 projects.}
}

\author{Authors omitted for double-blind review}

\author{\IEEEauthorblockN{
Luca Bertaccini\IEEEauthorrefmark{1},
Gianna Paulin\IEEEauthorrefmark{1},
Tim Fischer\IEEEauthorrefmark{1},
Stefan Mach\IEEEauthorrefmark{3},
Luca Benini\IEEEauthorrefmark{1}\IEEEauthorrefmark{2} \\
\IEEEauthorblockA{\IEEEauthorrefmark{1}IIS, ETH Zürich, Switzerland; \IEEEauthorrefmark{2}DEI, University of Bologna, Italy; \IEEEauthorrefmark{3}Axelera AI, Switzerland\\ \{lbertaccini, pauling, fischeti, lbenini\}@iis.ee.ethz.ch;
stefan.mach@axelera.ai} 
}}

\maketitle

\begin{abstract}  
Low-precision formats have recently driven major breakthroughs in neural network (NN) training and inference by reducing the memory footprint of the NN models and improving the energy efficiency of the underlying hardware architectures. Narrow integer data types have been vastly investigated for NN inference and have successfully been pushed to the extreme of ternary and binary representations. In contrast, most training-oriented platforms use at least 16-bit floating-point (FP) formats. Lower-precision data types such as 8-bit FP formats and mixed-precision techniques have only recently been explored in hardware implementations. We present MiniFloat-NN, a \riscv instruction set architecture extension for low-precision NN training, providing support for two 8-bit and two 16-bit FP formats and expanding operations. The extension includes sum-of-dot-product instructions that accumulate the result in a larger format and three-term additions in two variations: expanding and non-expanding. We implement an ExSdotp unit to efficiently support in hardware both instruction types. The fused nature of the ExSdotp module prevents precision losses generated by the non-associativity of two consecutive FP additions while saving around 30\% of the area and critical path compared to a cascade of two expanding fused multiply-add units. We replicate the ExSdotp module in a SIMD wrapper and integrate it into an open-source floating-point unit, which, coupled to an open-source \riscv core, lays the foundation for future scalable architectures targeting low-precision and mixed-precision NN training. A cluster containing eight extended cores sharing a scratchpad memory, implemented in 12\,nm FinFET technology, achieves up to 575 GFLOPS/W when computing FP8-to-FP16 GEMMs at 0.8\,V, 1.26\,GHz.

\end{abstract}

\section{Introduction}

With machine learning becoming ubiquitous, the demand for \ac{NN} training has increased exponentially. Today's \ac{NN} models require up to three orders of magnitude more total compute than only two years ago~\cite{liekeynote}. 
Therefore, to efficiently compute these workloads, hardware architectures for \ac{NN} training need to evolve very rapidly, especially since the required increase in performance and efficiency cannot be achieved by technology scaling alone and requires algorithmic and architectural improvements.
The exponential growth of machine learning, and the need to train \ac{NN} models, also translates into ever higher energy and carbon impacts~\cite{wu2022sustainable}, which further motivates investments toward cutting-edge energy-efficient architectures. 

More compact data types have recently driven fundamental breakthroughs for efficient \ac{NN} training and inference~\cite{rodriguez2018lower}. These new low-precision formats greatly improve the memory footprint of the \ac{NN} models and reduce the datapath size of the \acp{PE}, thereby improving the overall power consumption.
\ac{NN} algorithms have shown to be particularly resilient to the noise introduced by employing lower-precision formats. Especially, narrow integer data types for inference -- even as low as ternary and binary formats~\cite{alemdar2017ternary, qin2020binary} -- have been extensively researched. \ac{NN} training, however, demands the higher dynamic range provided by \ac{FP} formats and is usually performed using at least \mbox{$16$-bit} \ac{FP} data types. Only recently, \mbox{$8$-bit} formats have been explored in a mixed-precision setup with \mbox{$16$-bit} formats~\cite{wang2018training, sun2019hybrid}. 

Further performance and efficiency improvements can be achieved by matching algorithmic advancements in mixed, low-bitwidth training with microarchitectural enhancements, e.g., developing new functional units and accelerators. Most modern CPUs already support \ac{FMA} operations, which are more precise than performing separate \ac{FP} multiply and add instructions. 
As a fundamental operation for both inference and training of \ac{NN} models, the sum of dot products is another ideal candidate for
a dedicated hardware block.

Fusing multiple \acp{FMA} into a single sum of products -- or dot product -- unit allows for a single normalization and rounding step, thereby improving the area, timing, and power consumption of the functional unit.
Furthermore, this fusion can mitigate precision losses caused by the non-associativity of \ac{FP} additions. Even higher gains can be achieved by designing specialized matrix-multiplication accelerators in which dot-product modules are scaled out, as in the case of tensor cores~\cite{h100}. 
For \ac{NN}-oriented workloads, dot-product operations are particularly interesting when the accumulation result is returned in a wider format, which allows for retaining a higher computational precision. In this paper, we refer to such operations as \textit{expanding} operations.

Such a trend can already be observed in today's industry-leading architectures, where the A64FX~\cite{okazaki2020supercomputer}, the H100 GPU~\cite{h100}, and the IBM training accelerator presented in \cite{lee20217} provide three examples of systems with an increasing level of specialization. 
The first architecture is the most flexible and targets high-performance computing with a \mbox{$512$-bit} SIMD \ac{FPU} capable of vectorial \acp{FMA}. The GPU further specializes in data-parallel computation by reducing the control overhead even more than vector processors. It contains a large set of CUDA cores, capable of computing one \ac{FMA} instruction per cycle, and a number of large tensor cores, each capable of performing up to $1024$ dot-product operations per cycle with $8$-bit \ac{FP} precision.
The third design is fully specialized for efficient \ac{NN} training and inference. It implements two $8\times8$ arrays of mixed precision elements, each one supporting eight expanding dot products using $8$-bit \ac{FP} formats or eight $16$-bit \acp{FMA} per cycle.

Focusing on \riscv-based open-source academic designs, an interesting point in the design space is provided by Manticore~\cite{zaruba2020manticore}, a chiplet-based hierarchically-scalable architecture that builds upon the replication of clusters where eight compute and one DMA cores share a scratchpad memory.
Here, a tiny \riscv core~\cite{zaruba2020snitch} is coupled to a large \ac{FP} accelerator~\cite{mach2020fpnew} and enhanced with \ac{ISA} extensions that maximize the \ac{FPU} utilization. Manticore does not support the currently trending low-precision key \ac{FP} formats, nor a fused operation unit for the fundamental operation of \ac{NN} computation, i.e., the expanding sum of dot products.

In this paper, we present a novel parameterized \ac{FP} SIMD unit supporting the core operations of \ac{NN} training with support for two \mbox{8-bit} and two \mbox{16-bit} formats (\figref{fig:fpfmt}). New formats for the unit can be rapidly defined thanks to its easy parameterization scheme.
We evaluate this hardware unit standalone and in the context of an open-source \riscv core cluster derived from Manticore, where we integrate the novel \ac{NN} training capabilities. The main contributions of this work are the following:

\begin{enumerate}[leftmargin=*]
    \item We design an open-source parameterized multi-format unit supporting \ac{ExSdotp} instructions, as well as non-expanding and expanding three-operand additions, called \ac{Vsum} and \ac{ExVsum}. The hardware unit enables a $2\times$ speedup with respect to computing on \acp{ExFMA} while reducing the area and critical path by 30\% compared to a cascade of two \ac{ExFMA} units and preventing precision losses due to the non-associativity of two consecutive \ac{FP} additions. We integrate the \ac{ExSdotp} unit into an open-source multi-format \ac{FPU} called FPnew\footnote{\url{https://github.com/pulp-platform/fpnew}. ExSdotp unit currently available at \url{https://github.com/pulp-platform/fpnew/tree/feature/expanding_dotp}}~\cite{mach2020fpnew}. 
    \item We specify MiniFloat-NN: a \riscv \ac{ISA} extension for low-precision \ac{FP} training on many-core architectures exploiting the new computational units. We integrate the enhanced \ac{FPU} into an open-source \riscv $8$-core cluster based on Snitch cores\footnote{\url{https://github.com/pulp-platform/snitch}}~\cite{zaruba2020snitch} supporting the new \ac{ISA} extension and sharing a fast-access software-managed scratchpad memory.
    \item We carry out a detailed evaluation of the standalone \ac{ExSdotp} unit and the enhanced compute cluster, providing area, performance, energy efficiency, and accuracy results.
\end{enumerate}

\section{Related Work}

\subsection{Floating-Point Formats for NN Training}

\begin{figure}[t]
\centering
\includegraphics[width=.925\columnwidth]{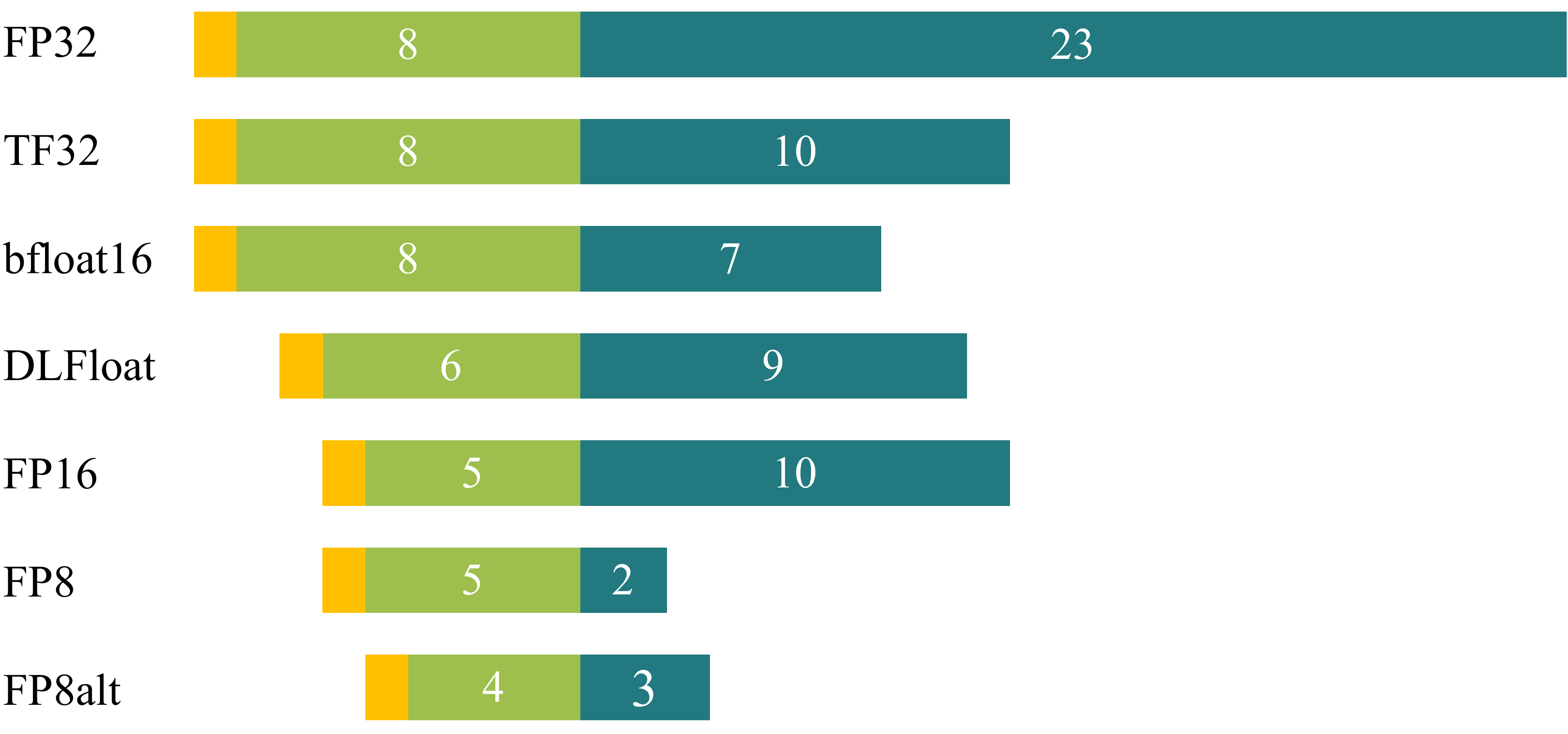}  
\caption{Relevant floating-point formats in the context of NN training. The exponent (green) and mantissa (blue) bitwidths are reported for each data type.}
\label{fig:fpfmt}
\vspace{-0.3cm}
\end{figure}

The shift towards less-than-32-bit formats for training -- which started with the \texttt{FP16} data type -- witnessed a quite large set of new proposed formats. An overview of the relevant \ac{FP} formats in the context of \ac{NN} training is provided in \figref{fig:fpfmt}. As many training algorithms benefit from a higher dynamic range than the one provided by \texttt{FP16}, \texttt{\ac{TF32}}~\cite{choquette2021nvidia} and \texttt{bfloat16}~\cite{bf16} recently gained traction. Both formats preserve the \mbox{$8$-bit} exponent and the corresponding dynamic range of \texttt{FP32} while reducing the number of mantissa bits to $10$ bits in the first case and $7$ bits in the second case. 

As the mantissa datapath dominates the area and power consumption of \ac{FP} engines, reducing the mantissa width allows placing more computational units at the same area cost and enables larger efficiency improvements than lowering the number of exponent bits. 
Additionally, \texttt{bfloat16} has been further optimized by handling rounding and special cases differently from the \mbox{IEEE-754} directives, e.g., by flushing subnormals to zero~\cite{bf16}.
On the A100 GPU, \texttt{\ac{TF32}} enabled a $10\times$ performance speedup with respect to \texttt{FP32} computations on its predecessor, the V100 GPU~\cite{choquette2021nvidia}, while using \texttt{bfloat16} on the NVIDIA A100 rather than \texttt{\ac{TF32}} enables a $2\times$ higher performance. However, due to the low number of mantissa bits, the accumulation of \texttt{bfloat16} products is usually performed in \texttt{FP32}. To mitigate the need for larger-precision accumulation, IBM introduced \texttt{DLFloat}~\cite{agrawal2019dlfloat}, a \mbox{$16$-bit} format composed of a $6$-bit exponent and a $9$-bit mantissa, thereby providing an intermediate dynamic range and precision with respect to \texttt{FP16} and \texttt{bfloat16}. Finally, as different applications benefit more from different data types, Nannarelli proposed a variable precision \mbox{$16$-bit} format~\cite{nannarelli2020variable} that can be set to represent \texttt{FP16}, \texttt{bfloat16}, \texttt{DLFloat}, and an additional data type with $7$-bit exponents and a $9$-bit exponent.

The significant benefits of using low-precision formats pushed researchers to investigate and demonstrate the feasibility of training models with \mbox{$8$-bit} \ac{FP} formats~\cite{wang2018training, sun2019hybrid}. Two formats, which we call \texttt{FP8} and \texttt{FP8alt}, have gathered particular interest. \texttt{FP8} consists of a $5$-bit exponent, thus providing the same dynamic range as \texttt{FP16}, and a $2$-bit mantissa, while \texttt{FP8alt} features a $4$-bit exponent and a $3$-bit mantissa. Sun \etal~\cite{sun2019hybrid} demonstrated considerable improvements for a large set of \ac{NN} training tasks by employing \texttt{FP8alt} in the forward propagation and \texttt{FP8} in the backward pass. However, these studies relied on software emulation and were not reproduced on hardware platforms. NVIDIA recently released the H100 GPU~\cite{h100}, where these \mbox{$8$-bit} formats are supported and provide a $2\times$ speedup with respect to \mbox{$16$-bit} data types. Nonetheless, hardware architectures supporting such formats are still rare and not well studied from the application viewpoint. 

The vast set of \ac{FP} data types being investigated and proposed motivates the need for a flexible open-source hardware platform in which new formats could be rapidly explored and supported.

\subsection{Related Architectures}

\begin{figure}[t]
\vspace{-0.2cm}
\centering
\includegraphics[width=\columnwidth]{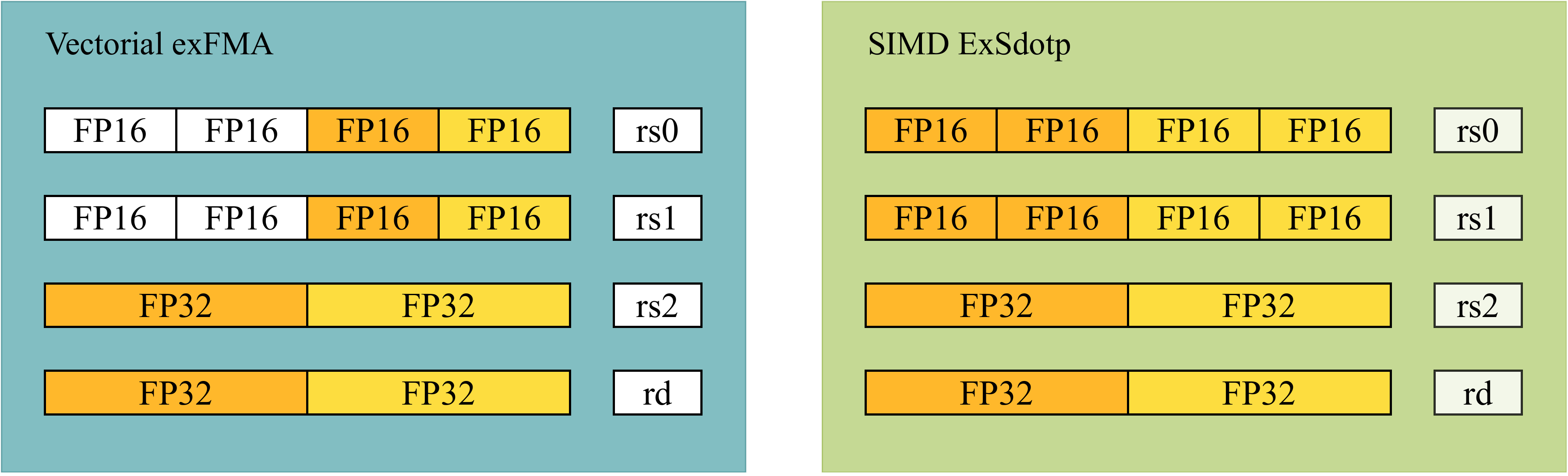}
\caption{Register file utilization: ExFMA vs. ExSdotp. In the ExFMA case, only half of two source registers can be processed each cycle, while using ExSdotp allows for fully exploiting the information saved in the register file,
and that can be passed to the FPU interface.
}
\label{fig:regfile}
\vspace{-0.2cm}
\end{figure}

As a critical operation for a wide set of kernels, most modern processors provide support for \ac{FMA} instructions, usually in a non-expanding fashion. However, in the context of \ac{NN} training, expanding operations get particularly interesting, as they allow using low-precision formats while retaining high accuracy~\cite{henry2019leveraging}. To address this need, various \acp{ExFMA} implementations have been investigated. Brunie~\cite{brunie2017modified} proposed an \acp{ExFMA} unit multiplying \texttt{FP16} inputs and accumulating in larger precision, while Mach \etal~\cite{mach2020fpnew} designed an \ac{FPU} for transprecision computing, containing a multi-format \ac{FMA} capable of computing \acp{ExFMA} on a wide set of \ac{FP} formats. 
However, \acp{ExFMA} do not use the \ac{FP} register file efficiently, as shown in the left part of \figref{fig:regfile}. Due to their unbalanced nature, they access only half of the data in two source registers, thus not fully exploiting all the information that can be packed into those registers while entirely using the third source register and the destination register. Furthermore, to use the entire register file space, sub-word accesses would be needed, as well as multiple instructions to cover all possible source locations. 
An \ac{ExSdotp} instruction would instead consume all the available data, as shown in the right half of \figref{fig:regfile}, preventing these drawbacks.

\begin{figure}[t]
\vspace{-0.2cm}
\centering
\includegraphics[width=\columnwidth]{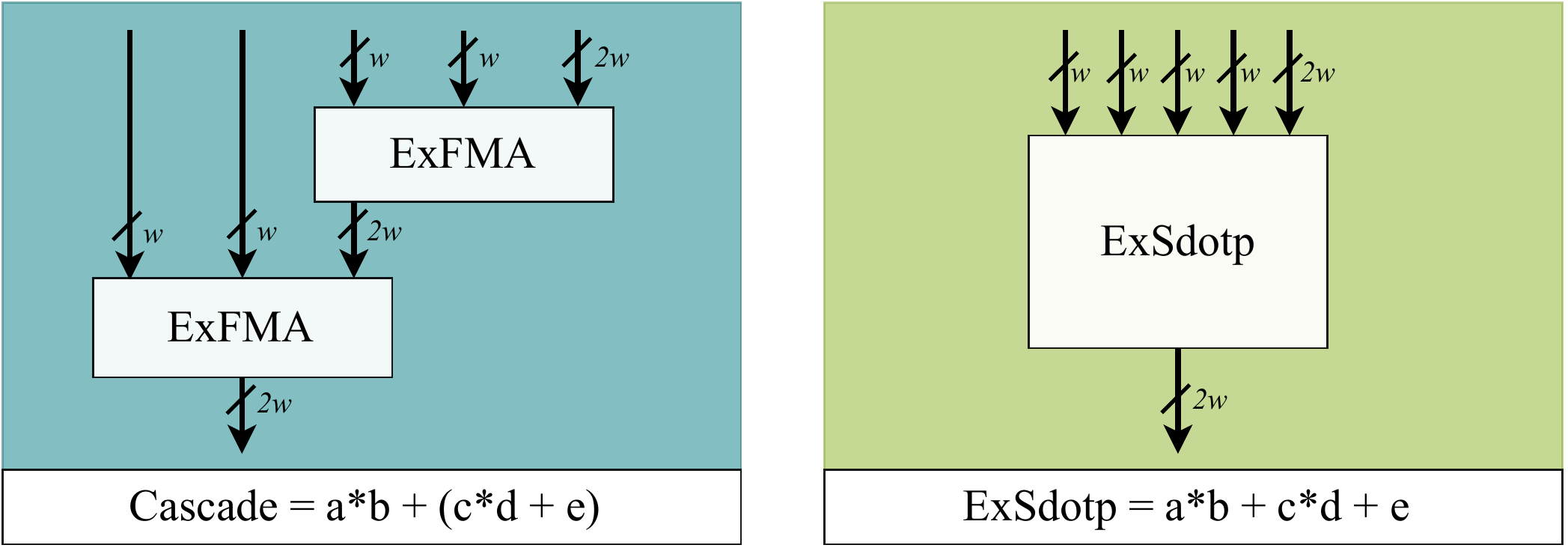}
\caption{ExSdotp instruction computed by two ExFMA units vs. one dedicated ExSdotp unit. Note that the first solution actually computes $a*b + (c*d + e)$, which is not necessarily equal to $a*b + c*d + e$ when using \ac{FP} arithmetic.}
\label{fig:formulas}
\vspace{-0.2cm}
\end{figure}

An expanding dot-product unit with accumulation computing \mbox{$a\times b + c\times d + e$} can be designed in a discrete or fused fashion. The first one places two consecutive \ac{ExFMA} modules in a cascade, as shown in \figref{fig:formulas}. A fused design requires more engineering effort, as it involves implementing the non-trivial three-term \ac{FP} addition~\cite{tao2012three}. However, as a fused design allows for a single normalization and rounding step, improving the module's area, timing and accuracy, it is often the best choice. 

Sohn and Swartzlander~\cite{sohn2013improved} developed a fused \ac{FP} dot-product unit computing  \mbox{$a\times b + c\times d$}. Additionally, they designed a non-expanding three-term \ac{FP} adder~\cite{sohn2014fused}, which, merged with their dot-product unit, can generate a dot-product unit with accumulation.
These two designs were only implemented for single precision and double precision, and were not designed for expanding operations. Intel designed a fused floating-point many-term dot-product unit~\cite{hickmann2020intel} computing $32$ products and accumulating them in higher precision for the Nervana \ac{NN} processor.
The unit computes the products in \texttt{bfloat16} and accumulates the result in single precision. However, it does not support \mbox{8-bit} formats, which are now getting traction in the context of \ac{NN} training. The Nervana dot-product unit increases its internal datapath to reduce, but not fully prevent, precision losses when cancellation causes a large normalization shift. Such losses would have a higher impact on low-precision \ac{FP} formats, as their final results are more sensitive to small variations.

Zhang \etal~\cite{zhang2019efficient} and Mao \etal~\cite{mao2021configurable} proposed two academic designs with multiple-precision dot-product capabilities. However, also these modules do not support \mbox{$8$-bit} \ac{FP} data types. On the contrary, the IBM AI chip~\cite{lee20217} supports \texttt{FP8} and \texttt{FP8alt} expanding dot products but is only capable of \texttt{DLFloat} \ac{FMA}. 

As low-precision data types present a high tradeoff between accuracy and dynamic range, supporting multiple formats allows for dynamically adapting to the application requirements. The \ac{FPU} developed by Nannarelli~\cite{nannarelli2020variable} addresses this by defining a new variable-precision data type that can be set to \texttt{FP16}, \texttt{bfloat16}, \texttt{DLFloat}, and a fourth \mbox{$16$-bit} custom format. However, no support for \mbox{$8$-bit} formats was considered.

Finally, FPnew~\cite{mach2020fpnew}, being highly parameterized, can work on a wide set of \ac{FP} data types, from \mbox{$8$-bit} to \mbox{$64$-bit}, and allows for a fast definition of new formats. Nonetheless, it does not provide support for dot-product instructions. Due to its highly configurable environment, we took the open-source FPnew as a starting point for our work. FPnew also provides a second advantage: it is included in a flexible and highly efficient compute cluster~\cite{zaruba2020snitch} that can be hierarchically replicated to form a large many-core system~\cite{zaruba2020manticore}, enabling a fast exploration of our extensions on a real hardware platform.

In this work, we develop an open-source \ac{ExSdotp} unit working on two \mbox{$8$-bit} and two \mbox{$16$-bit} formats and accumulating in higher precision, $16$ bits and $32$ bits, respectively. Our design is highly parameterized so that new formats can be easily defined and explored.
We integrate a SIMD \ac{ExSdotp} module in a lightweight open-source \riscv processor called Snitch~\cite{zaruba2020snitch} to create a compute cluster with low-precision \ac{NN}-training capabilities and extend its \ac{ISA} with a custom \riscv extension that we named \textit{Minifloat-NN}.

\vspace{-0.1cm}
\section{Architecture}
\vspace{-0.1cm}
In the following, we discuss the supported \ac{FP} formats in \secref{subsec:formats} and describe the architecture of our new \ac{FP} \ac{ExSdotp} unit capable of computing \ac{ExVsum} and \ac{Vsum} on the same datapath in \secref{subsec:exsdotp} and \ref{subsec:vsum}. The unit's integration into an open-source modular energy-efficient multi-format \ac{FPU} is discussed in \secref{subsec:simd}, while the architecture of our evaluation \ac{PE} with the MiniFloat-NN extension and using our enhanced \ac{FPU} with SIMD \ac{ExSdotp} capabilities is introduced in \secref{subsec:pe}.

\subsection{Supported FP Formats}\label{subsec:formats}
The high parametrization of the open-source FPnew unit allows not only to select a specific set of supported \ac{FP} formats but also to quickly define new formats, thereby enabling fast research on new data types. 
For this work, we add the \texttt{FP8alt} definition and enable the following \ac{FP} formats:

\begin{table}[]
\vspace{-0.2cm}
\caption{Combinations of source and destination formats supported by the ExSdotp unit for each operation}
\label{comb_fmts}
\resizebox{\columnwidth}{!}{%
\begin{tabular}{@{}lccccc@{}}
\cmidrule(l){2-6}
                 & \multicolumn{5}{c}{\textbf{Destination}}                       \\ \cmidrule(l){2-6} 
\textbf{Source} & \texttt{FP32} & \texttt{FP16alt} & \texttt{FP16} & \texttt{FP8} & \texttt{FP8alt} \\ \midrule
\texttt{FP32}    & Vsum           & -              & -              & -    & -    \\
\texttt{FP16alt} & ExSdotp/ExVsum & Vsum           & Vsum           & -    & -    \\
\texttt{FP16}    & ExSdotp/ExVsum & Vsum           & Vsum           & -    & -    \\
\texttt{FP8}     & -              & ExSdotp/ExVsum & ExSdotp/ExVsum & Vsum & Vsum \\
\texttt{FP8alt}  & -              & ExSdotp/ExVsum & ExSdotp/ExVsum & Vsum & Vsum \\ \bottomrule
\end{tabular}%
}
\vspace{-0.25cm}
\end{table}

\begin{itemize}
    \item \texttt{FP64}: 11-bit exponent, 52-bit mantissa
    \item \texttt{FP32}: 8-bit exponent, 23-bit mantissa
    \item \texttt{FP16}: 5-bit exponent, 10-bit mantissa
    \item \texttt{FP16alt}: 8-bit exponent, 7-bit mantissa
    \item \texttt{FP8}: 5-bit exponent, 2-bit mantissa
    \item \texttt{FP8alt}: 4-bit exponent, 3-bit mantissa
\end{itemize}

\texttt{FP16alt} matches the exponent and mantissa widths of widely-used \texttt{bfloat16} but follows the IEEE-754 directives for rounding and subnormal number handling.

The extended \ac{FPU} supports all the \riscv \ac{FP} instructions, except division and square root, for all the enabled formats.
For our novel \ac{ExSdotp}-capable unit, we focus on low-precision formats, where precision loss prevention is most crucial. 
The expanding operations compute from $8$ to \mbox{$16$-bit} and from $16$ to $32$-bit formats, while the non-expanding \ac{Vsum} is implemented for $8$, $16$, and $32$-bit \ac{FP} formats, as summarized in \tabref{comb_fmts}.

\subsection{ExSdotp Unit} \label{subsec:exsdotp}
The \ac{ExSdotp} unit takes five operands, four inputs expressed in a $w$-bit source format (\textit{src\_format}), and an accumulator input in a $2w$-bit destination format (\textit{dst\_format}) to compute a sum of dot products in the $2w$-bit destination format:
\begin{equation}
    ExSdotp_{2w} = a_w \times b_w + c_w \times d_w + e_{2w}
\end{equation}

\begin{figure}[t]
\vspace{-0.2cm}
\centering
\includegraphics[width=\columnwidth]{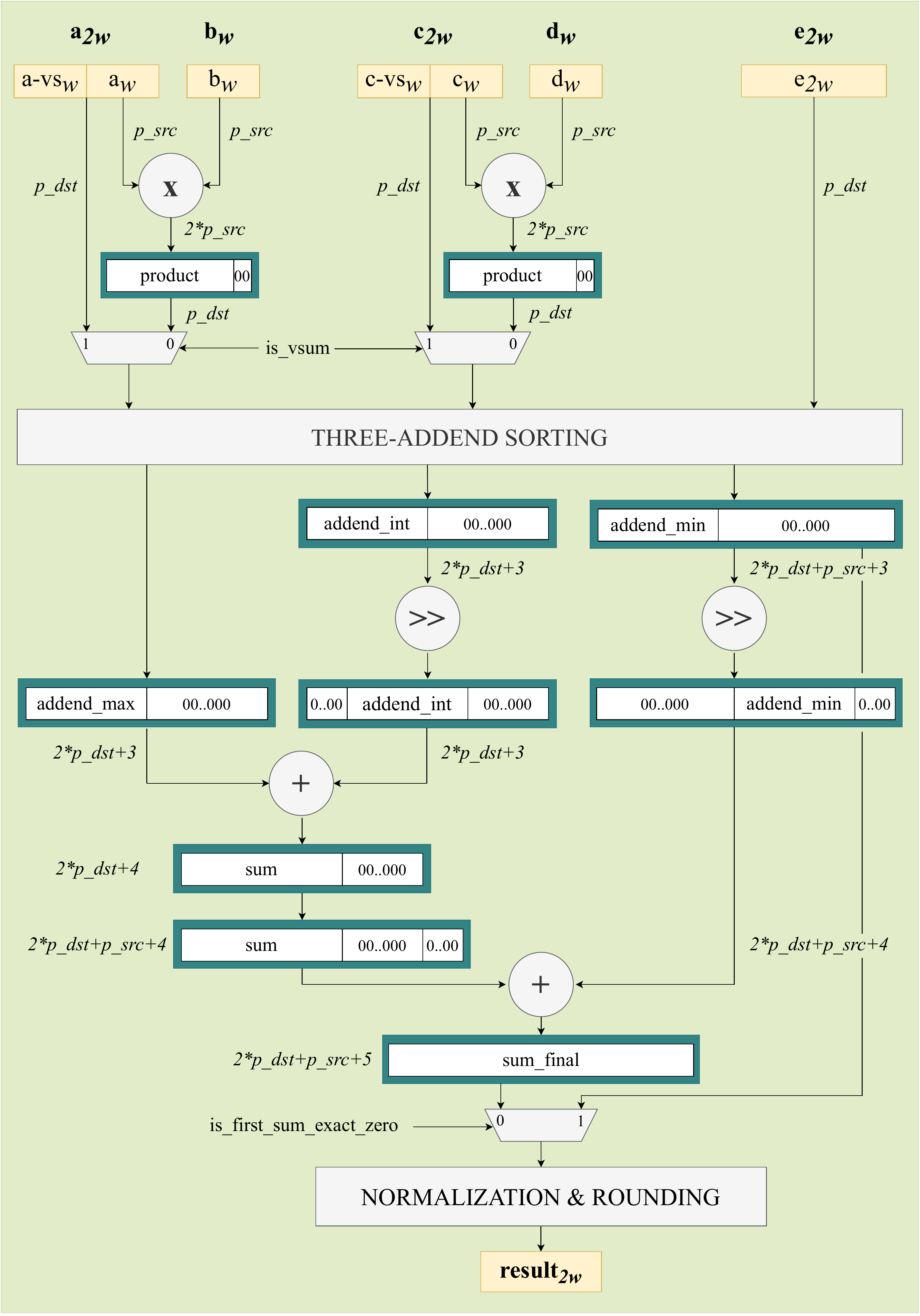}
\caption{ExSdotp data flow: examples of how the information is packed at each stage of the datapath are provided in the blue boxes (note that we do not include the exponent datapath in this figure, nor the sticky bits that are generated after the various shifts). The ExSdotp operation takes four $w$-bit inputs and a $2w$-bit accumulator, while the Vsum takes three $2w$-bit inputs to produce a $2w$-bit output.}
\label{fig:exdotp_datapath}
\vspace{-0.375cm}
\end{figure}

\begin{figure*}[t]
\vspace{-0.3cm}
\centering
\includegraphics[width=0.9\textwidth]{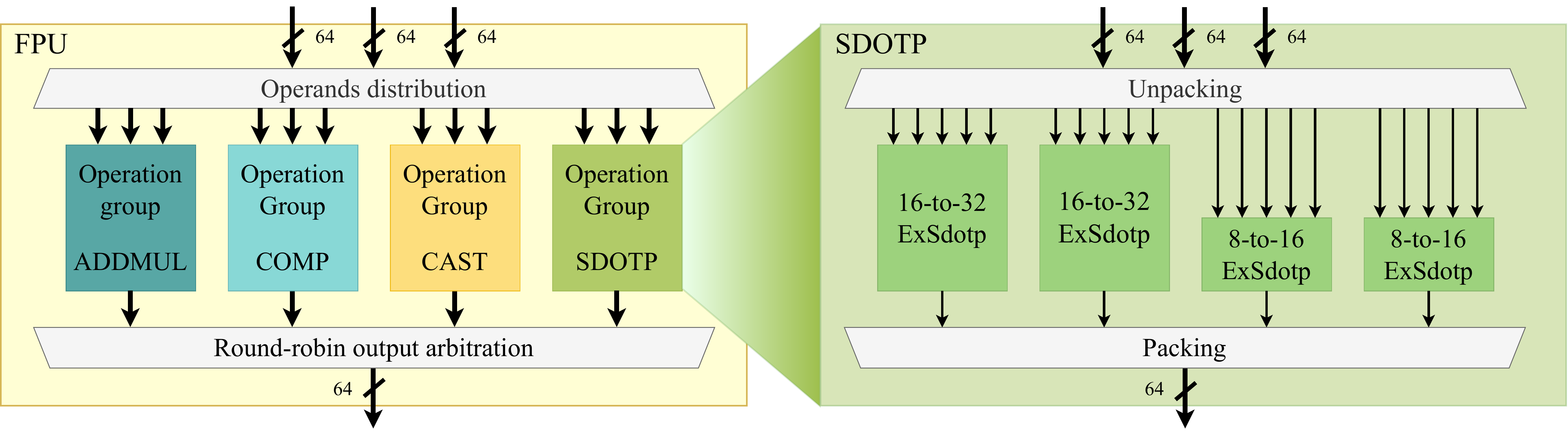}
\caption{Block diagram of the extended FPU, with a zoom on the ExSdotp SIMD module.}
\label{fig:ExSDDotp SIMD}
\vspace{-0.3cm}
\end{figure*}

The \ac{ExSdotp} module handles subnormals as all the other IEEE-754 operations 
and supports a parametric number of pipeline stages.
Each instance of the \ac{ExSdotp} unit is constrained by the largest exponent and mantissa widths enabled by the source and destination format parameterization. 
Enabled formats with narrower exponent and/or mantissa fields are mapped to the lower and upper bits of the wider exponent and mantissa field, respectively. 
This mapping scheme allows performing lower-precision computations on the same datapath, as well as adding new slightly different \ac{FP} formats at a very low area overhead.
A $16$-to-$32$-bit module can support all the format combinations specified in \tabref{comb_fmts}, while a narrower $8$-to-\mbox{$16$-bit} unit can support all the combinations for \ac{ExSdotp} but \{\textit{src\_format:} \texttt{FP16}, \textit{dst\_format:} \texttt{FP32}\}, and \{\textit{src\_format:} \texttt{FP16alt}, \textit{dst\_format:} \texttt{FP32}\}.

The data flow in the \ac{ExSdotp} unit is shown in \figref{fig:exdotp_datapath}. For the sake of simplicity, the diagram does not depict the exponent datapath, which is responsible for computing the exponent differences used to sort the two products and the accumulator and to calculate the shift amounts. The maximum number of mantissa bits plus one (the hidden one) is called $p\_src$ and $p\_dst$ for source and destination formats, respectively, and indicates the precision of the floating-point formats.

Initially, the two mantissa products are computed, producing $(2\times p\_src)$-wide results. Note that $(2\times p\_src)$ differs from $p\_dst$; e.g., for an \texttt{FP16}-to-\texttt{FP32} \ac{ExSdotp}, the former is $22$ bits, and the latter is $24$ bits.
To match the same precision as the accumulator $e_{p\_dst}$ the mantissa products are padded with zeroes to $\text{prod\_ab}_{(p\_dst)}$
and $\text{prod\_cd}_{(p\_dst)}$ of width $p\_dst$.
\begin{align}
\text{prod\_ab}_{(p\_dst)} &= \{a_{(p\_src)}\times b_{(p\_src)}, &0_{(p\_dst-2\times p\_src)} \} 
\end{align}

After that step, what is left is a three-term addition. Three-term fused \ac{FP} additions present additional challenges due to \ac{FP} additions being non-associative~\cite{tao2012three}. Two consecutive \ac{FP} additions might produce different results if performed in different orders; for example if \mbox{$|a| >> |c|$} and \mbox{$b = -a$}, then \mbox{$(a + b) + c = c$}; however, \mbox{$a + (b + c)$} might return $0$, as, if $c$ is small enough, \mbox{$(b + c)$} will result in $b$. 

To cope with these challenges, we sort the three addends, finding the maximum $\text{max}_{(p\_dst)}$, the intermediate $\text{int}_{(p\_dst)}$, and the minimum absolute value $\text{min}_{(p\_dst)}$. After being zero-padded to match the increasing internal precision (e.g., $\{\text{min}_{(p\_dst)}, 0_{(p\_dst+3)}\}$), the intermediate and the minimum addends are right-shifted by their exponent difference to the maximum addends $(\text{exp\_max} - \text{exp\_min})$.
\begin{equation}\label{eq:sum}
\text{sum}_{(2\times p\_dst+4)} = \text{max}_{(2\times p\_dst+3)} + \text{int}_{(2\times p\_dst+3)}
\end{equation}

After summing the maximum and intermediate addends \eqref{eq:sum}, an additional \textit{p\_src} bits are added by zero-padding ($\{\text{sum}_{(2\times p\_dst+4)}, 0_{p\_src}\}$) to prevent catastrophic cancellations when the maximum addend is the result of a product between a normal and a subnormal value.
In the last step before the normalization and rounding, the minimum addend is accumulated to the padded $\text{sum}$.
\begin{equation}
\begin{aligned}
\text{sum\_final}_{(2\times p\_dst+p\_src +5)} = &\text{sum\_max}_{(2\times p\_dst+p\_src+4)} \\
&+ \text{min}_{(2\times p\_dst+p\_src+4)}
\end{aligned}
\end{equation}
Summing the addends with the largest absolute values first and gradually increasing the bitwidths at each step allows us to prevent precision losses that could occur when performing two \ac{FP} additions.
If the first sum produces a non-zero value, the increased precision will ensure enough precision bits even in case of cancellation in the second addition.
Else, if the first sum produces an exact zero result, the possibly useful shifted-out bits of the minimum addend are recovered by directly assigning the minimum addend to the result of the second sum.

\subsection{ExVsum and Vsum on the ExSdotp Datapath}\label{subsec:vsum}
By setting the inputs $b_w$ and $d_w$ to one, the \ac{ExSdotp} unit can easily perform an \ac{ExVsum} \eqref{eq:exvsum}.
\begin{equation}\label{eq:exvsum}
    ExVsum_{2w} = a_w + c_{w} + e_{2w}
\end{equation}
As discussed in \secref{subsec:exsdotp}, the unit already contains all the logic necessary to perform a non-expanding three-term addition. To enable non-expanding \ac{Vsum} \eqref{eq:vsum} in the larger \textit{dst\_fmt}, we increase the size of two operand inputs from the \textit{src\_format} width to the \textit{dst\_format} width by extending operand \textit{a} and \textit{c} with the \textit{a\_vs} and \textit{c\_vs} fields. The support for \ac{Vsum} is added by bypassing the two mantissa multiplications, as shown in \figref{fig:exdotp_datapath}. 
\begin{equation}\label{eq:vsum}
    Vsum_{2w} = a_{2w} + c_{2w} + e_{2w}
\end{equation}
Such an operation can be used to reduce and accumulate the results packed in a register after SIMD \ac{ExSdotp} executions (see \figref{fig:regfile}).

\subsection{SIMD Wrapper and Integration into FPnew}\label{subsec:simd}

\begin{figure}[t]
\vspace{-0.1cm}
\centering
\includegraphics[width=\columnwidth]{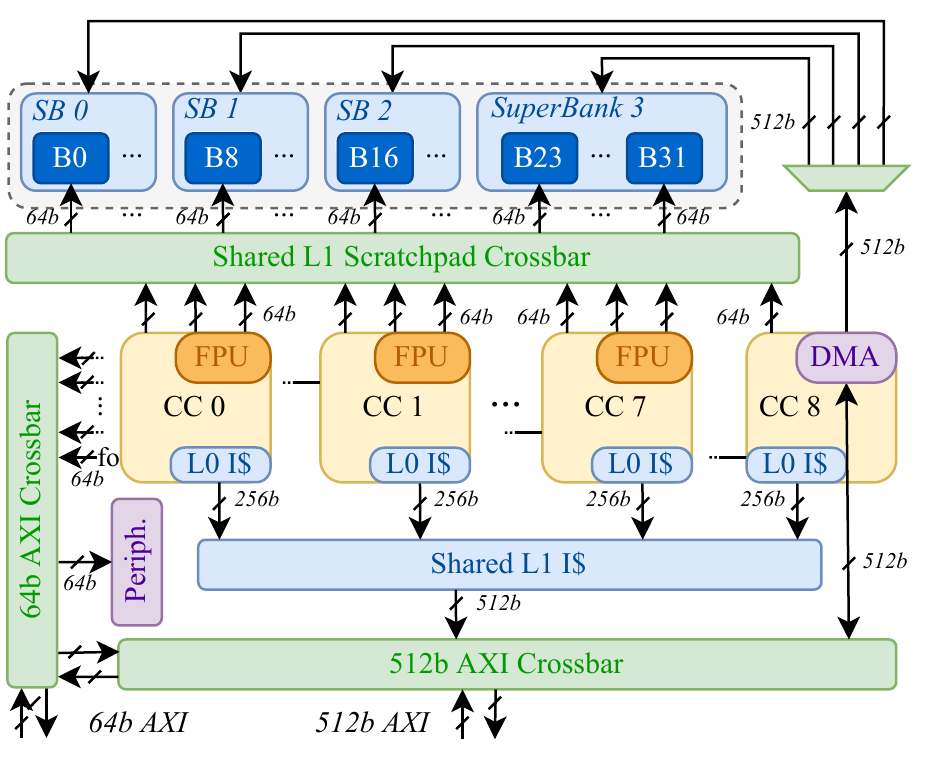}
\vspace{-0.6cm}
\caption{A cluster composed of eight Snitch compute cores, each one coupled with FPnew extended with ExSdotp capabilities, plus an extra DMA core.}
\label{fig:cluster}
\vspace{-0.2cm}
\end{figure}

FPnew is natively organized in modules, each one responsible for one operation group: ADDMUL, DIVSQRT, COMP (comparison), and CONV (conversion). When instantiating the top level, each module can be deactivated through a parameter. The configuration of the \ac{FPU} coupled to Snitch has the DIVSQRT disabled, thus not containing the correspondent module. For our evaluation \ac{PE}, we integrate an \ac{ExSdotp} SIMD wrapper into FPnew as a new operation group module, SDOTP. Since the proposed \ac{PE} supports double-precision instructions, the \ac{FP} register file contains \mbox{$64$-bit} entries. That allows for packing two \texttt{FP32}, four \texttt{FP16}/\texttt{FP16alt}, or eight \texttt{FP8}/\texttt{FP8alt} values in a single \ac{FP} register. The FPnew interface accepts up to three $64$-bit input operands and produces one \mbox{$64$-bit} output per cycle. Therefore, we organized our SIMD wrapper in two \mbox{$16$-to-$32$-bit} and two \mbox{$8$-to-$16$-bit} \ac{ExSdotp}, which means it will compute up to two $1$6-to-$32$-bit or four \mbox{$8$-to-$16$-bit} \ac{ExSdotp} operations each cycle. The SIMD wrapper is also responsible for unpacking the five operands from the \mbox{$64$-bit} input and packing the result into $64$-bits, as shown in \figref{fig:ExSDDotp SIMD}.

\subsection{MiniFloat-NN PE}\label{subsec:pe}

We build our evaluation \ac{PE} upon Snitch~\cite{zaruba2020snitch}, a tiny open-source $32$-bit \riscv processor coupled with an FPnew~\cite{mach2020fpnew} instance supporting single and double-precision \ac{FP} through a latency-tolerant acceleration interface. Snitch uses two custom \ac{ISA} extensions to enable an \ac{FPU} utilization of above $90$\%. The \ac{SSR} extension maps a regular load or store access pattern to fixed floating-point registers, effectively eliminating most of the implicit load and store instructions; while the \ac{FREP} extension allows to buffer and repeat a sequence of \ac{FPU} instructions to prevent the loop overhead caused by branching instructions, and to relieve pressure on the instruction cache.

\begin{figure}[t]
\vspace{-0.2cm}
  \begin{minipage}{0.5\columnwidth}
    \centering
    \includegraphics[width=\columnwidth]{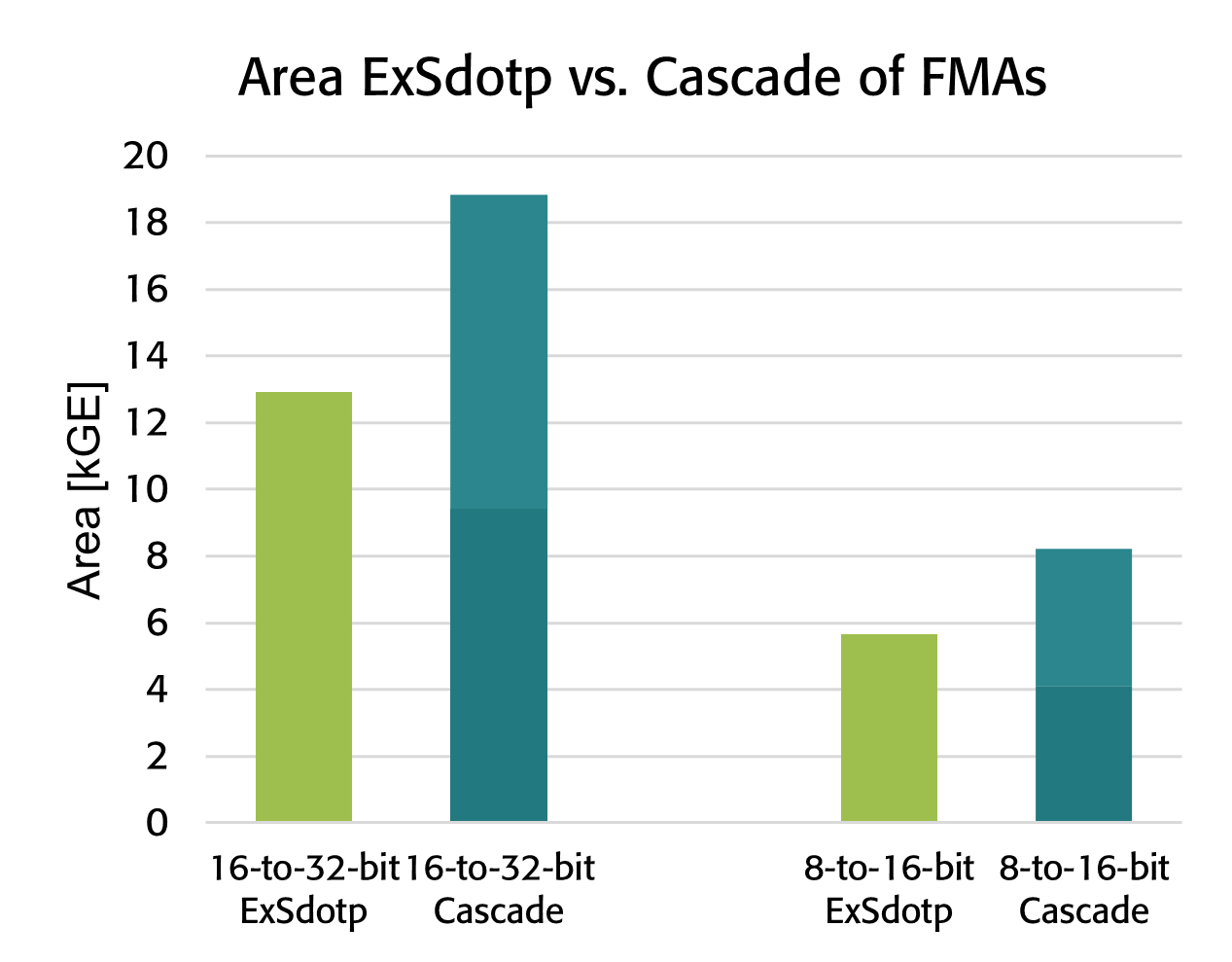}
    \subcaption{}
    \label{fig:area_comparison}
  \end{minipage}\hfill%
  \begin{minipage}{0.5\columnwidth}
    \centering
    \includegraphics[width=\columnwidth]{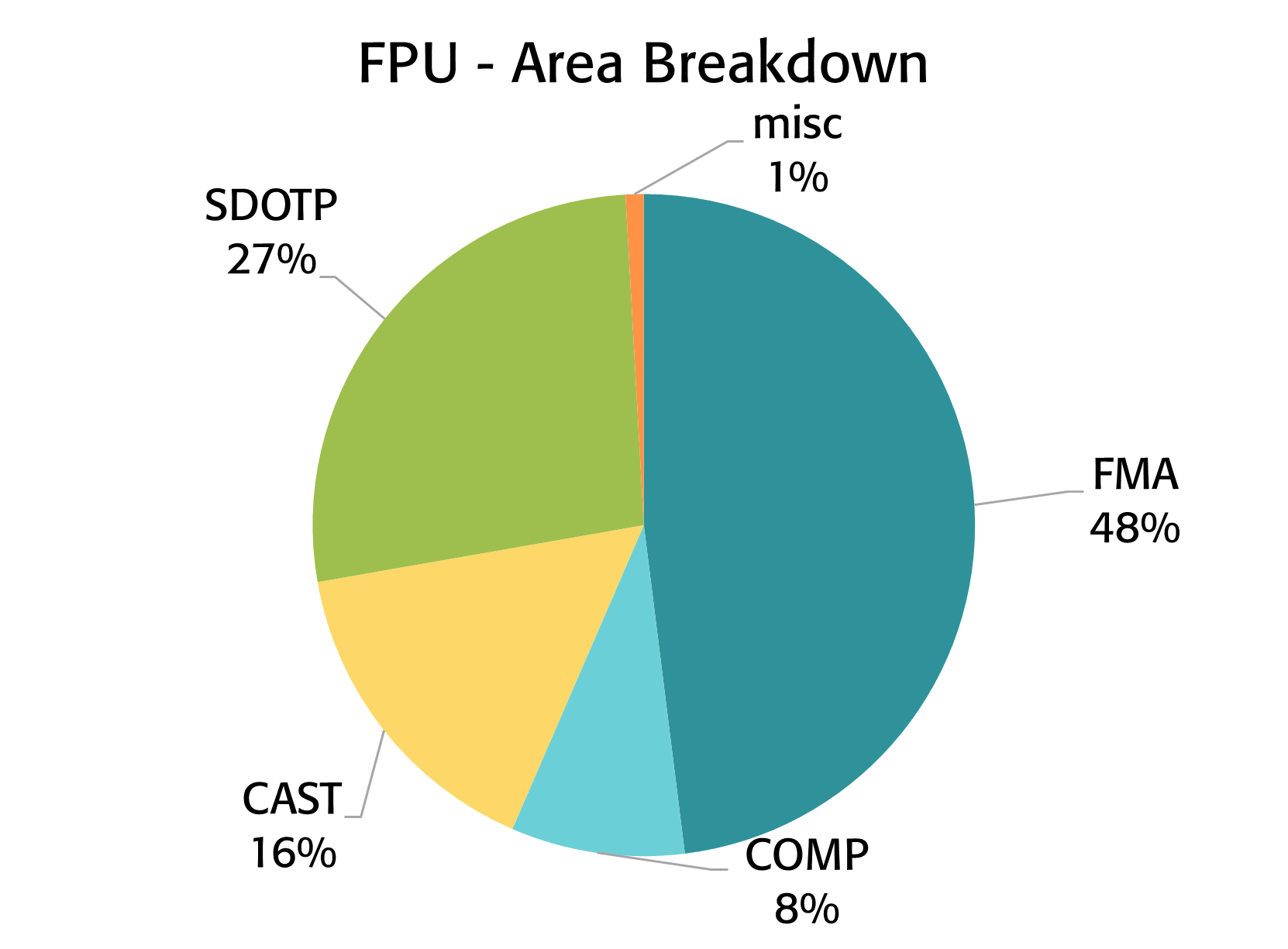}
    \subcaption{}
    \label{fig:area_break}
  \end{minipage}\hfill%

  \caption{Area results: (a) Comparison of a $16$-to-$32$-bit and an $8$-to-$16$-bit ExSdotp units against a set of two ExFMA modules that arranged in a cascade can perform a non-fused expanding-sum-of-dot-product instruction; (b) Area breakdown of the extended FPU.}
  \label{fig:fp_toplevel}
  \vspace{-0.2cm}
\end{figure}

To build the MiniFloat-NN \ac{PE}, we replaced the native FPnew instance with our enhanced \ac{FPU} and extended the Snitch decoder to support the new instructions. The levels of pipeline registers were set to $3$ for the SDOTP operation group, $3$ for the ADDMUL, $2$ for CAST, and $1$ for COMP. Due to the limited encoding space, we did not replicate the same instruction for different \ac{FP} formats sharing the same width. Instead, the alternative formats -- \texttt{FP16alt} and \texttt{FP8alt} -- are controlled by two additional bits, $src\_is\_alt$ and $dst\_is\_alt$, in the \ac{FP} \ac{CSR}. An \texttt{FP16alt} kernel will then differ from an \texttt{FP16} kernel by a single \ac{CSR} write.
The MiniFloat-NN extension augments the smallFloat\footnote{https://iis-git.ee.ethz.ch/smach/smallFloat-spec} extension by adding \ac{ExSdotp}, \ac{ExVsum}, and \ac{Vsum} SIMD instructions:
\begin{itemize}
    \item \texttt{exsdotp} \texttt{rd, rs1, rs2}
    \item \texttt{exvsum} \hspace{0.09cm} \texttt{rd, rs1, rs2}
    \item \texttt{vsum} \hspace{0.52cm} \texttt{rd, rs1}
\end{itemize}
For all the three instruction types \texttt{rd} also behaves as \texttt{rs3/rs2}, being the accumulator, and contains packed data in higher precision than the ones packed in \texttt{rs1/rs2}.

To evaluate our extensions, we replace the native Snitch core with the proposed MiniFloat-NN-capable \ac{PE} in a Snitch cluster, where a set of eight such compute \acp{PE} share a $32$-bank scratchpad memory, a DMA core, and an L1 instruction cache, as shown in \figref{fig:cluster}. 

\section{Experimental Results}
We evaluate the proposed \ac{ExSdotp} unit and the enhanced compute cluster for area, performance, energy efficiency, and accuracy, and compare them against related \ac{SoA} architectures.

\begin{table}[]
\vspace{-0.2cm}
\centering
\caption{Performance of the MiniFloat-NN-capable Snitch cluster\textsuperscript{\dag}}
\label{tabperf}
\centering
\resizebox{\columnwidth}{!}{%
\begin{threeparttable}
\begin{tabular}{@{}cccccc@{}}
\cmidrule(l){2-6}
 & \multicolumn{3}{c}{\textbf{FMA-based}} & \multicolumn{2}{c}{\textbf{ExSdotp-based}} \\ \cmidrule(l){2-6} 
\textbf{GEMM size} & \textbf{\begin{tabular}[c]{@{}c@{}}\texttt{FP64}\\ {$[$}\si{\cycles}{$]$}\end{tabular}} & \textbf{\begin{tabular}[c]{@{}c@{}}\texttt{FP32}\\ {$[$}\si{\cycles}{$]$}\end{tabular}} & \textbf{\begin{tabular}[c]{@{}c@{}}\texttt{FP16}\textsuperscript{*}\\ {$[$}\si{\cycles}{$]$}\end{tabular}} & \textbf{\begin{tabular}[c]{@{}c@{}}\texttt{FP16} to \texttt{FP32}\textsuperscript{*}\\ {$[$}\si{\cycles}{$]$}\end{tabular}} & \textbf{\begin{tabular}[c]{@{}c@{}}\texttt{FP8} to \texttt{FP16}\textsuperscript{*}\\ {$[$}\si{\cycles}{$]$}\end{tabular}} \\ \midrule
\textbf{$64\times64$} & \cellcolor[HTML]{FFFFFF}$37306$ & \cellcolor[HTML]{FFFFFF}$20195$ & \cellcolor[HTML]{FFFFFF}$12232$ & \cellcolor[HTML]{FFFFFF}$10968$ & \cellcolor[HTML]{FFFFFF}$7019$ \\
\textbf{$64\times128$} & \cellcolor[HTML]{FFFFFF}- & \cellcolor[HTML]{FFFFFF}$38058$ & \cellcolor[HTML]{FFFFFF}$20726$ & \cellcolor[HTML]{FFFFFF}$20169$ & \cellcolor[HTML]{FFFFFF}$11165$ \\
\textbf{$128\times128$} & \cellcolor[HTML]{FFFFFF}- & \cellcolor[HTML]{FFFFFF}- & \cellcolor[HTML]{FFFFFF}$83890$ & \cellcolor[HTML]{FFFFFF}$80709$ & \cellcolor[HTML]{FFFFFF}$43244$ \\
\textbf{$128\times256$} & \cellcolor[HTML]{FFFFFF}- & \cellcolor[HTML]{FFFFFF}- & \cellcolor[HTML]{FFFFFF}- & \cellcolor[HTML]{FFFFFF}- & \cellcolor[HTML]{FFFFFF}$82501$ \\ \bottomrule \\
\end{tabular}%
\begin{tablenotes}[]
    \centering
    \small
    \vspace{-0.2cm}
    \item [\dag] Only GEMM sizes for which all the data can fit in the local memory.
    \item [*] The results for the alt-format kernels are not reported as they differentiate from the standard-format kernels by a single CSR write.
\end{tablenotes}
\end{threeparttable}}
\vspace{-0.3cm}
\end{table}

\begin{table*}[] 
\vspace{-0.25cm}
\caption{Comparison of FPUs supporting low-precision formats (top four rows) and Evaluation of the MiniFloat-NN extension (bottom two rows)}
\label{tab:soa}
\resizebox{\textwidth}{!}{%
\begin{threeparttable}
\begin{tabular}{@{}lcccccccccccc@{}}
\toprule
\multirow{2}{*}{\textbf{Design}} & \multirow{2}{*}{\textbf{Technology}} & \multirow{2}{*}{\textbf{Voltage}} & \multirow{2}{*}{\textbf{Frequency}} & \multirow{2}{*}{\textbf{Area}} & \multirow{2}{*}{\textbf{DotP}} & \multicolumn{4}{c}{\textbf{ Performance $[$\si{\flop\per\cycle}$]$\tnote{a}}} & \multirow{2}{*}{\textbf{\begin{tabular}[c]{@{}c@{}}\vspace{-0.05cm}Peak\\Throughput\\{$[$}\si{\giga\flops}{$]$}\end{tabular}}} & \multirow{2}{*}{\textbf{\begin{tabular}[c]{@{}c@{}}Efficiency\tnote{d}\\ {$[$}\si{\giga\flops\per\watt}{$]$}\end{tabular}}} \\ \cmidrule(lr){7-10}
 &  &  &  &  &  & \texttt{FP16alt}\tnote{c} & \texttt{FP16}\tnote{c} & \texttt{FP8}\tnote{c} & \texttt{FP8alt}\tnote{c} &  &  \\ \midrule
\textbf{ExSdotp FPU} & \SI{12}{\nano\meter} & \SI{0.8}{\volt} & \SI{1.26}{\giga\hertz} & \SI{0.019}{\square\milli\meter} & yes & $8/8$ & $8/8$ & $16/16$ & $16/16$ & $20.2$ (\texttt{exFP8}) & $1631$  (\texttt{exFP8}) \\
\textbf{FPnew~\cite{mach2020fpnew}{}} & \SI{22}{\nano\meter} & \SI{0.8}{\volt} & \SI{0.923}{\giga\hertz} & \SI{0.049}{\square\milli\meter} & no & $4/8$ & $4/8$ & $8/16$ & -$/$- & $14.8$ (\texttt{FP8}) & $1245$ (\texttt{FP8}) \\
\textbf{Mao \etal~\cite{mao2021configurable}\tnote{b}} & \SI{28}{\nano\meter} & \SI{1.0}{\volt} & \SI{1.43}{\giga\hertz} & \SI{0.013}{\square\milli\meter} & yes & -$/$- & -$/20$ & -$/$- & -$/$- & $28.6$ (\texttt{FP16}) & $975$ (\texttt{FP16}) \\
\textbf{Zhang \etal~\cite{zhang2019efficient}\tnote{b}} & \SI{90}{\nano\meter} & \SI{1.0}{\volt} & \SI{0.667}{\giga\hertz} & \SI{0.191}{\square\milli\meter} & yes & -$/$- & $8/8$ & -$/$- & -$/$- & $5.3$ (\texttt{FP16}) & $113$ (\texttt{FP16}) \\ \midrule
\textbf{MiniFloat-NN Snitch} & \SI{12}{\nano\meter} & \SI{0.8}{\volt} & \SI{1.26}{\giga\hertz} & \SI{0.52}{\square\milli\meter} & yes & $8/8$ & $8/8$ & $16/16$ & $16/16$ & 160 (\texttt{exFP8}) & $575$ (\texttt{exFP8}) \\
\textbf{Snitch~\cite{zaruba2020snitch}} & \SI{22}{\nano\meter} & \SI{0.8}{\volt} & \SI{1}{\giga\hertz} & \SI{0.66}{\square\milli\meter} & no & -$/$- & -$/$- & -$/$- & -$/$- & $16$ (\texttt{FP64}) & $80$ (\texttt{FP64})\tnote{e} \\ \bottomrule 
\end{tabular}%
\begin{tablenotes}[]
    \centering
    \small
    \vspace{0.2cm}
    \item [a] We report only the performance related to low-precision formats. All the designs in the table support \texttt{FP32} and \texttt{FP64} as well.
    \item [b] Only FMA and ExSdotp unit. No support for other operation groups as cast and comparison.
    \item [c] Performance reported in the form \textit{expanding}$/$\textit{non-expanding}, and considering $1$ ExSdotp = $4$ FLOP, and $1$ FMA = $2$ FLOP.
    \item [d] Peak efficiency for the FPUs; efficiency achieved computing GEMMs for the clusters.
    \item [e] The Snitch cluster in \cite{zaruba2020snitch} obtained \SI{104}{\giga\flops\per\watt} for \texttt{FP32} GEMMs. As it was evaluated without vectorial \ac{FMA} support, the increased efficiency derives from the reduced activity when performing \texttt{FP32} \acp{FMA} on an \texttt{FP64} \ac{FMA} datapath.

\end{tablenotes}
\end{threeparttable}}
\vspace{-0.15cm}
\end{table*}

\subsection{Area and Timing}
We use \textsc{Synopsys Fusion Compiler 2021.06} to synthesize the \ac{ExSdotp} unit and synthesize, place, and route the Minifloat-NN-capable cluster in \textsc{GlobalFoundries}' \SI{12}{\nano\meter} FinFET technology. First, we consider the \ac{ExSdotp} unit standalone, without any pipeline stage, and target a relaxed clock period of \SI{333}{\mega\hertz} in a worst-case corner (\SI{0.72}{\volt}, \SI{125}{\celsius}). Since two \ac{ExFMA} modules, arranged in a cascade, can perform a dot product, we compare our $16$-to-$32$-bit and $8$-to-\mbox{$16$-bit} units against a set of two \ac{ExFMA} modules supporting the same data types. Note that the cascade of \ac{ExFMA} units will not compute exactly the same operation as it rounds twice and does not mitigate potential adverse effects of the non-associativity of \ac{FP} additions. Furthermore, a cascade of \ac{ExFMA} cannot compute \ac{Vsum}. Since the second \ac{ExFMA} unit in the cascade requires the output of the first as an input, in an implementation with no pipeline registers, each \ac{FMA} instance is required to work at \SI{667}{\mega\hertz} to ensure that the cascade will run at \SI{333}{\mega\hertz} as the \ac{ExSdotp} unit. We synthesize the \ac{ExFMA} unit with such a constraint and provide the area results in \figref{fig:area_comparison}. The \ac{ExSdotp} occupies around $30\%$ less area than two \acp{ExFMA} and shows around $30\%$ of critical path reduction.

\begin{figure}[t]
\centering
\includegraphics[width=\columnwidth]{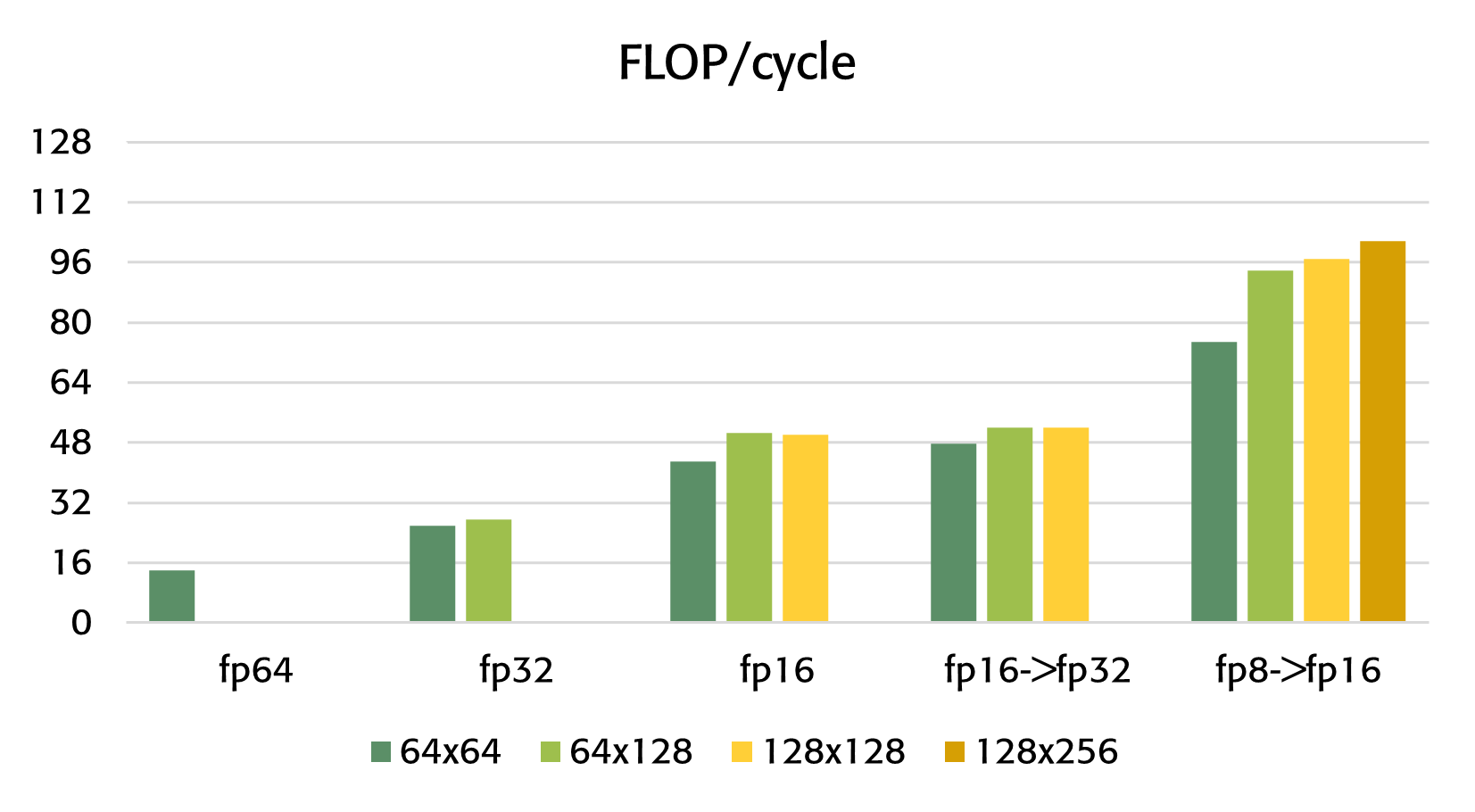}
\vspace{-0.7cm}
\caption{Performance: \si{\flop\per\cycle} per \ac{FP} format for different  GEMM sizes.} 
\label{fig:mac}
\vspace{-0.25cm}
\end{figure}

Then, we synthesize, place and route the extended Snitch cluster, targeting \SI{950}{\mega\hertz} in the worst-case scenario. To achieve the higher frequency, we insert in the enhanced \ac{FPU} $3$ levels of pipeline registers for the SDOTP operation group, $3$ for the ADDMUL, $2$ for the CAST, and $1$ for the COMP. The extended computing cluster occupies \SI{4.3}{\mega\GE} and, in a typical corner (\SI{0.8}{\volt}, \SI{25}{\celsius}), achieves \SI{1.26}{\giga\hertz}. The \ac{ExSdotp} SIMD module occupies \SI{44.5}{\kilo\GE}, amounting to $27\%$ of the overall area of the \ac{FPU}, which occupies \SI{165}{\kilo\GE} (\figref{fig:area_break}).

\subsection{Performance}

\begin{figure}[t]
\vspace{-0.2cm}
\centering
\includegraphics[width=\columnwidth]{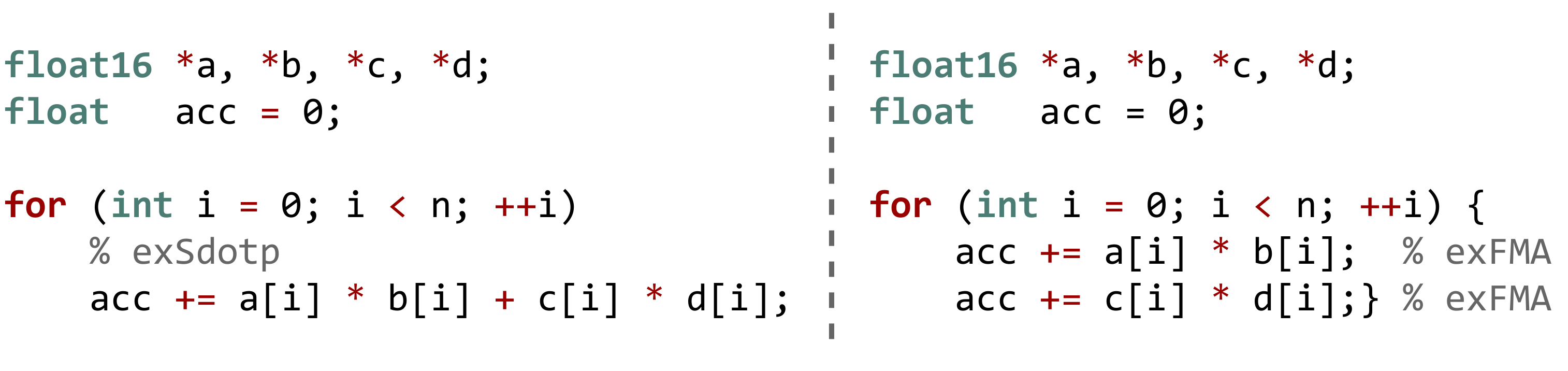}
\caption{ExSdotp and ExFMA-based accumulation of products. The result gets accumulated in higher precision. The figure shows the \texttt{FP16}-to-\texttt{FP32} case.}
\label{fig:prec}
\vspace{-0.25cm}
\end{figure}

We implement and measure a collection of \ac{FMA} and \ac{ExSdotp}-based \ac{GEMM} kernels for different formats and problem sizes. We consider only \ac{GEMM} sizes that fit entirely into the \SI{128}{\kilo\byte} local scratchpad memory.  The kernels are compiled with an extended LLVM-12 compiler using intrinsics for our new instructions. All implemented kernels make use of the \ac{SSR} and \ac{FREP} custom \ac{ISA} extensions of Snitch~\cite{zaruba2020snitch}.
We run the kernels on the enhanced Snitch cluster and measure the execution cycles in a cycle-accurate RTL simulation using \textsc{Mentor Questasim}.

Using our new expanding instructions allows reducing the memory footprint remarkably.
We can fit up to $128\times256$ \acp{GEMM} into the memory when leveraging the new $8$-to-$16$-bit \ac{ExSdotp} support, while \texttt{FP16}-only and \texttt{FP64} kernels fit a size of $128\times128$ and $64\times64$, respectively.

The MiniFloat-NN extension enables a peak utilization of $8$ \si{\flop\per\cycle} for $16$-to-$32$-bit \acp{GEMM}, and $16$ \si{\flop\per\cycle} for $8$-to-$16$-bit \acp{GEMM}. That doubles the peak performance achievable with \acp{ExFMA} whose SIMD implementation suffers from inefficient register file usage, see \figref{fig:regfile}. Moreover, it doubles the peak utilization compared with non-expanding \acp{FMA} computing in the \textit{dst\_format}, and matches the peak performance for the non-expanding \acp{FMA} computing in the \textit{src\_format} while providing higher internal precision.

The total execution cycles and achieved performance are reported in \tabref{tabperf} and \figref{fig:mac}, respectively. 
Using a $16$-to-$32$-bit \ac{ExSdotp} reduces the number of cycles by up to $10\%$ compared to the \texttt{FP16} \ac{FMA} kernel.
This benefit comes from the expanding operation, which halves the number of intermediate results that get reduced at the end of the computation compared to the \ac{FMA}-based kernel.

With a constant problem size, the speed-up when halving the format size is, in the worst case, $1.56\times$ instead of the ideal $2\times$. The reason for this deviation is the overhead generated by setting up the \ac{SSR} stream and initializing registers, which stays largely the same while the number of compute instructions inside the repeated loop is halved. Such overheads impact less and less on the results when increasing the problem size.
The $128\times256$ \texttt{FP8}-to-\texttt{FP16} \ac{GEMM} achieves $1.96\times$  and $7.23\times$ the \si{\flop\per\cycle} of the $128\times128$ \texttt{FP16}-to-\texttt{FP32} \ac{GEMM} and the $64\times64$ \texttt{FP64} \ac{GEMM}, respectively. Note that the \texttt{FP64} kernel, as non-expanding and non-SIMD, does not suffer from these additional overheads.

\subsection{Power and Energy Efficiency}

\begin{table}[t]
\vspace{-0.1cm}
\centering
\caption{Accuracy comparison between ExSdotp and ExFMA}
\label{tab:accu}
\resizebox{\columnwidth}{!}{%
\begin{threeparttable}
\begin{tabular}{@{}lcccccccc@{}}
\toprule
\multirow{2}{*}{\textbf{Operation}} & & \multirow{2}{*}{\textbf{Format}} & & \multicolumn{5}{c}{\textbf{\begin{tabular}[c]{@{}c@{}}Relative error vs. \texttt{FP64}\textsuperscript{\dag} \end{tabular}}} \\ \cmidrule(l){5-9} 
\multicolumn{1}{l}{} & \multicolumn{1}{c}{} & & & $n = 500$ & & $n = 1000$ & & $n = 2000$ \\ \midrule
\textbf{ExSdotp} & & \texttt{FP16}-to-\texttt{FP32} & & $0$ & & $1.1 \times 10\textsuperscript{-7}$ & & $5.4 \times 10\textsuperscript{-7}$ \\
\textbf{ExFMA} & & \texttt{FP16}-to-\texttt{FP32} & & $7.6 \times 10\textsuperscript{-7}$ & & $1.8 \times 10\textsuperscript{-6}$ & &  $9.9 \times 10\textsuperscript{-7}$ \\ \midrule
\textbf{ExSdotp} & & \texttt{FP8}-to-\texttt{FP16} & & $5.9 \times 10\textsuperscript{-4}$ & & $2.7 \times 10\textsuperscript{-3}$ & & $3.9 \times 10\textsuperscript{-3}$  \\
\textbf{ExFMA} & & \texttt{FP8}-to-\texttt{FP16} & & $5.9 \times 10\textsuperscript{-4}$ & & $8.2 \times 10\textsuperscript{-3}$ & & $1.2 \times 10\textsuperscript{-2}$  \\ \bottomrule 
\end{tabular}%
\begin{tablenotes}[]
    \centering
    \small
    \vspace{0.1cm}
    \item [\dag] The golden \texttt{FP64} result is converted to \texttt{FP32}/\texttt{FP16} for the error calculation.
\end{tablenotes}
\end{threeparttable}
}
\vspace{-0.3cm}
\end{table}
To assess the power consumption and energy efficiency of the placed-and-routed extended cluster, we simulate an \ac{ExSdotp}-based \ac{GEMM} with \textsc{Mentor Questasim}, annotating the switching activity data. We extract the average power consumption with \textsc{Synopsys PrimePower} under typical conditions (\SI{0.8}{\volt}, \SI{25}{\celsius}). When computing $128\times256$ \mbox{\texttt{FP8}-to-\texttt{FP16}} \acp{GEMM} at \SI{1.26}{\giga\hertz}, the MiniFloat-NN cluster achieves \textbf{\SI{128}{\giga\flops}} consuming \SI{224}{\milli\watt}, thus achieving \SI{575}{\giga\flops\per\watt} (where $1$ \ac{ExSdotp} is counted as \SI{4}{\flop}). 
Our extended cluster compares favorably with the native Snitch system, which, implemented in a 22\,nm technology (at \SI{1}{\giga\hertz}, \SI{0.8}{\volt}, \SI{25}{\celsius}) in \cite{zaruba2020snitch}, reaches \SI{80}{\giga\flops\per\watt} when computing \texttt{FP64} \acp{GEMM}.
Our extended cluster working on \mbox{\texttt{FP8}-to-\texttt{FP16}} kernels reaches $7.2\times$ the efficiency of the native Snitch system computing in double precision.
We summarize the benefits introduced at a system level by our \mbox{MiniFloat-NN} \riscv extension by comparing the extended cluster against its baseline version in the bottom two rows of \tabref{tab:soa}.

\subsection{Accuracy}
To evaluate the accuracy of the proposed \ac{ExSdotp} unit, we accumulate an increasing number of dot products. We generate the inputs randomly, with a Gaussian distribution, in the source precision. We then perform the accumulations using: \textit{(i)} low-precision \acp{ExSdotp}, \textit{(ii)} low-precision \acp{ExFMA}, and \textit{(iii)} \texttt{FP64} \acp{ExFMA}. The first two implementations are shown in \figref{fig:prec}, while we report the comparison result of the \ac{ExSdotp} unit against the \ac{ExFMA} in terms of relative error against the \texttt{FP64} golden model in \tabref{tab:accu}.

As different errors can compensate during the accumulation, the precision results vary with the selected number of inputs. However, the \ac{ExSdotp} unit consistently shows better accuracy than the \ac{ExFMA} for \texttt{FP16}-to-\texttt{FP32} and \texttt{FP8}-to-\texttt{FP16} workloads. The absolute accuracy improvement grows when the input operands have smaller bitwidths.

\subsection{SoA comparison}
We compare our enhanced \ac{FPU} against its baseline counterpart, FPnew, and two recent \ac{SoA} architectures with low-precision support in \tabref{tab:soa}. 
The \acp{PE} proposed by Mao \etal~\cite{mao2021configurable} and Zhang \etal~\cite{zhang2019efficient} compute SIMD \ac{FMA} or dot-product operations with different precisions, \mbox{$16$}-bit and higher, while FPnew also supports \texttt{FP8}. However, none of them can work with \texttt{FP8alt} data.
Our extended \ac{FPU}, thanks to the proposed \ac{ExSdotp} SIMD unit, achieves the highest energy efficiency among the selected mixed-precision \acp{FPU}. It outperforms the module developed by Zhang \etal~\cite{zhang2019efficient} by $14.4\times$, and the multiple-precision \ac{PE} by Mao \etal~\cite{mao2021configurable} by $1.7\times$. 
Furthermore, it provides $30\%$ higher efficiency than FPnew working with \texttt{FP8} data and doubles its peak performance when using expanding operations. 

\section{Conclusion}

We presented an \ac{ISA} extension for low-precision \ac{NN} training on \riscv cores. The new instructions are carried out on the proposed SIMD unit composed of a set of modules computing expanding \ac{FP} dot products and reusing the same hardware to calculate expanding and non-expanding three-term additions. The \ac{ExSdotp} unit supports two $16$-bit and two $8$-bit input formats. Thanks to the module's parameterization, new formats can be rapidly defined and explored.
The proposed \ac{ExSdotp} module performs twice the computations and exploits the \ac{FP} register file more efficiently than an expanding \ac{FMA} while providing higher accuracy. The \ac{ISA} extension has finally been implemented in an open-source \ac{PE}, composed of a tiny \riscv processor coupled with an \ac{FPU} enhanced with our SIMD \ac{ExSdotp} unit.
A cluster containing eight of such extended \acp{PE} implemented in 12\,nm technology achieves \SI{160}{\giga\flops} of peak performance and \SI{575}{\giga\flops\per\watt} of energy efficiency when computing \texttt{FP8}-to-\texttt{FP16} GEMMs at \SI{1.26}{\giga\hertz}, \SI{0.8}{\volt}.
\vspace{-0.1cm}

\Urlmuskip=0mu plus 1mu\relax
\bibliographystyle{IEEEtran}
\bibliography{bibl.bib}

\end{document}